\documentclass[aip,jmp,reprint]{revtex4-1}
\usepackage{amssymb,amsmath,latexsym,amsthm}

\begin{document}

\title[Quantum uncertainty and holomorphic maps]{The uncertainty principle and the energy identity for
holomorphic maps in geometric quantum mechanics}

\author{Barbara A. Sanborn}
\email{bsanborn@antiochcollege.edu.}
\affiliation{Antioch College, Yellow Springs, OH 45387}

\def\pf{\noindent {\bf Proof:} }
\def\endpf{\phantom\opensquare\hfill\llap{\opensquare}\medskip}
\def\opensquare{\hbox{$\rlap{$\sqcap$}\sqcup$}}
\def\calM{\mathcal{M}}
\def\calH{\mathcal{H}}
\def\calE{\mathcal{E}}
\def\frakg{\mathfrak{g}}

\newtheorem{theorem}{Theorem}
\newtheorem{prop}[theorem]{Proposition}
\newtheorem{lemma}[theorem]{Lemma}
\newtheorem{cor}[theorem]{Corollary}
\newtheorem{definition}[theorem]{Definition}

\begin{abstract}
 The theory of geometric quantum mechanics describes a quantum system as
a Hamiltonian dynamical system with a complex projective Hilbert
space as its phase space, thus equipped with a Riemannian metric in addition to a symplectic structure.
This paper extends the geometric quantum theory to include aspects of the symplectic topology of the
state space by identifying the Robertson-Schr\"{o}dinger uncertainty relation for pure quantum states
as the differential version of the energy identity in the theory of $J$-holomorphic curves.
We consider a family of maps from a Riemann surface into a finite-dimensional quantum state space 
by using the vector fields generated by two quantum observables,
 and show that the Fubini-Study metric tensor pulls back 
 by such a map to the covariance tensor for the two observables. By calculating
 the map energy density in the pull-back metric,
the uncertainty relation is represented as an equality that compares the map energy differential to
the sum of the pull-back of the symplectic form and a positive definite term 
that vanishes when the map is holomorphic. 
Saturation of the Robertson-Schr\"{o}dinger inequality occurs when the map is conformal and
the off-diagonal covariance terms vanish. 
For compact Riemann surfaces where such a map can be globally defined, if the map is holomorphic, 
it is harmonic and its image is a minimal surface.
In this case, the uncertainty product integral 
is a topological invariant that depends only on the homology class of the curve modulo its boundary.

\end{abstract}

\maketitle

\section{Introduction}

By describing a quantum system as a Hamiltonian dynamical system, geometric quantum mechanics 
emphasizes the symplectic geometry of the quantum state space
in a way similar to the geometric formulation\cite{Arnold,MR} of classical mechanics.
A distinctive feature of the quantum phase space is that its symplectic structure plays 
not only a dynamical role, but also determines the curvature of a connection on the 
canonical $U(1)$ bundle over the space, and plays a topological role as the 
Chern form of the associated complex line bundle.
The present work continues the development of this description of quantum mechanics as a 
Hamiltonian system while considering the global structure of the quantum state manifold, with the goal of
forging a link to work\cite{MS,HZ} in symplectic topology that has explained deep relationships 
between dynamical, geometric, and topological symplectic invariants.

The foundations of geometric quantum mechanics were laid
in the work of Chernoff and Marsden,\cite{CM} and
Kibble.\cite{Kibble} The theory has been developed by numerous
authors\cite{AS, BH, Boya, BSS, Cirelli90, Heslot, Hughston, MR, PV, RRG,
Schilling, Spera} into a full description of quantum mechanics as a Hamiltonian dynamical system on a
symplectic manifold. Technical issues raised for infinite dimensional phase spaces have been 
dealt with carefully.\cite{CM, Cirelli90} In the geometric theory,
the phase space of pure states of a quantum system is $P(\calH)$,
the projective space of a complex separable Hilbert space, $\calH$.
Schr\"{o}dinger dynamics on $P(\mathcal{H})$ is
the quantum version of Hamilton's equations, determined by the
natural symplectic structure on $P(\mathcal{H})$ induced by the imaginary part of the 
Hermitian inner product on $\calH$. 
 The real part of the inner product on $\calH$ induces the Fubini-Study metric on $P(\calH)$.
This additional Riemannian structure can be viewed as a source of features in
quantum systems that are distinctly different from classical
systems. In particular, the uncertainty in the expectation value of a
linear operator on $\calH$ is a measure of distance in this 
metric on $P(\calH)$.\cite{Anan90, AS, BH, Cirelli90, Hughston, PV, Schilling}
 A key result of geometric quantum mechanics expresses the quantum uncertainty
principle in terms of the symplectic form and Riemannian metric on 
$P(\calH)$.\cite{AS, BH, Cirelli90, Schilling} 
A geometric interpretation of the uncertainty principle has been given in terms of quantum Fisher 
information.\cite{Braunstein, Facchi, Gibilisco}

Further understanding of the symplectic geometry of quantum mechanics
came with a realization of the significance of Lie group actions and bundle structures
on the quantum phase space. The Hermitian inner product on $\calH$ determines a natural connection on 
the principal bundle $U(1)\hookrightarrow S(\calH)\rightarrow P(\calH)$, where
$S(\calH)=\{\psi \in \calH : |\psi|^{2}=1\}$. Since $P(\calH)$ is a homogeneous, isotropic manifold,
the curvature of this connection and the Chern form
on the associated line bundle are each scalar multiples of the pull-back of the 
symplectic form $\Omega$ on $P(\calH)$.
Thus, the holonomy, or geometric phase, of a closed path in $P(\calH)$
measures the symplectic area of a surface spanned by the path,\cite{Anan90, Boya, FA, Page}
and has been recognized\cite{BH} as the quantum version of the Poincar\'e integral invariant
which characterizes Hamiltonian systems.\cite{Arnold,HZ}
In applications restricting to a closed submanifold $\mathcal{N}$ of $P(\calH)$, $\Omega$ 
represents the first Chern class in $H^2(\mathcal{N}, \mathbb{Z})$; integrating $\Omega$  over
$\mathcal{N}$ yields the Chern number  
of the corresponding complex line bundle over $\mathcal{N}$. 

During the same period of time that symplectic structures
have come to be recognized in quantum physics, symplectic topology 
has emerged as a new field of mathematics devoted to the study of the global structure of
 symplectic manifolds.\cite{MS} A foundational
result is Gromov's\cite{Gromov} nonsqueezing theorem, which states
that if there is an embedding of the closed symplectic Euclidean
ball $B^{2n}(r)$ of radius $r$ into the symplectic cylinder
$B^{2}(R) \times \mathbb{R}^{2n-2}$ that preserves the symplectic
form, then $r \le R$. This result shows that there is a two-dimensional
quantity associated with symplectic manifolds, which is invariant
under symplectomorphisms. The symplectic capacities\cite{HZ, MS} are
defined to describe this symplectic area invariance. New methods
have been developed to study the group of symplectomorphisms of a
symplectic manifold, and its relation to the group of
volume-preserving diffeomorphisms. 
De Gosson and Luef\cite{deG, deG07} have related
symplectic capacitance and Gromov's non-squeezing theorem to 
the quantum covariance matrix and the uncertainty principle.

An especially powerful and useful tool for investigating the global structure
of symplectic manifolds is the theory of $J$-holomorphic
curves (or pseudoholomorphic curves) introduced by Gromov.\cite{Gromov}
Such curves generalize holomorphic maps between a Riemann surface and a complex manifold 
by relaxing the condition on the target manifold 
that its almost complex structure be integrable.\cite{Donaldson, McS04} 
The theory can also be useful for studying K\"{a}hler manifolds, particularly
for determining whether properties of holomorphic curves on these manifolds persist
when complex structure is perturbed. Our interest in applying insights from symplectic topology
to the study of quantum systems led us to
the central result of the present work, Theorem 7 of Section 3, which states 
that the Robertson-Schr\"{o}dinger uncertainty relation in quantum mechanics
is an example of the differential version of the energy identity in the theory of $J$-holomorphic 
curves. The early work on this result was originally published as part of the author's dissertation\cite{Sanborn}

The energy identity is related to a variational principle used to study smooth maps 
$\phi: \mathcal{N} \rightarrow \mathcal{M}$ between closed
Riemannian manifolds.\cite{EellsRatto, Eells}
Harmonic maps are the critical points of the energy functional,
\begin{equation}
E(\phi):=\frac{1}{2}\int_{\mathcal{N}}|d\phi|^{2}\,dV,
\end{equation}
where $|d\phi|$ \ denotes the Hilbert-Schmidt norm of the differential $d\phi$, 
regarded as a one-form on $\mathcal{N}$ with values in $\phi^{*}T(\calM)$,
and $dV$ denotes the volume element of the metric on $\mathcal{N}$. 
A useful relationship exists between the energy and volume functionals of $\phi$.
Minimal immersions are critical points of the volume functional,
\begin{equation}
V(\phi):=\int_{\mathcal{N}}|\Lambda ^n d\phi|\,dV,
\end{equation} 
where $\Lambda ^n d\phi$ denotes the Jacobian of $\phi$, and
$dV$ is now the volume element associated with the induced metric $\phi^{*}(g)$, where $g$ is the metric on $\calM$. 
Two results\cite{EellsRatto, Eells} from the variational theory are relevant to our study.  
1) If $\phi: \mathcal{N} \rightarrow \calM$  is a Riemannian (isometric) immersion, 
then $E(\phi)$ and $V(\phi)$ have the same critical points, 
that is, $\phi$ is harmonic if and only if $V(\phi)$ is stationary. 2) If dim($\mathcal{N})$=2, 
then for all smooth maps $\phi: \mathcal{N} \rightarrow \calM$, we have
$V(\phi) \le E(\phi)$, with equality holding if and only if $\phi$ is almost conformal; 
thus, if $\phi$ is harmonic and conformal, then $\phi$ minimizes the area functional and
$V(\phi) = E(\phi)$.

Harmonic maps have been useful as models for physical theories characterized 
by broken gauge symmetry,\cite{Misner} especially
Yang-Mills fields\cite{Bourguignon} and non-linear $\sigma$ models,\cite{Urakawa}
as well as in studies of minimal submanifolds.\cite{EellsRatto,Lawson}
Holomorphic maps between K\"{a}hler manifolds are a special case of harmonic maps between
Riemannian manifolds, and this case applies to our study of the quantum uncertainty principle.

The theory of $J$-holomorphic curves\cite{McS04,Audin} 
deals with smooth maps between almost complex manifolds,
\[u:(\Sigma,j)\rightarrow (\calM,J)\,,\]
where $\Sigma$ is a Riemann surface with complex structure $j$,
and the target manifold $\calM$ is  
equipped with a symplectic $\omega$ and almost complex structure $J$. 
Such a map $u$ is called $J$-holomorphic if it satisfies the generalized Cauchy-Riemann equation,
\begin{equation}
J \circ du = du \circ j. \label{JHolomorphicCondition}
\end{equation}
If $J$ is assumed to be $\omega$-compatible,
then there exists a unique Riemannian metric $g_J$ on $\calM$ defined as 
$(g_J)_x (\xi, \zeta)=\omega_x (\xi, J \zeta)$
for $\xi, \zeta \in T_x(\calM)$. In this case, if $\Sigma$ is compact,
the energy functional of the map $u$ is 
\begin{equation}
E(u)=\frac{1}{2} \int_{\Sigma} |du|_{g_{J}}^{2} dA_{\Sigma}\,,\label{Energy}
\end{equation}
where $dA_{\Sigma}$ is the area form on $\Sigma$.
The energy identity states that\cite{McS04}
\begin{equation}
E(u)=\int_{\Sigma}|\bar{\partial}_{J}(u)|_{g_{J}}^{2}\,dA_{\Sigma}+\int_{\Sigma} u^{*}\omega,
\label{energyequality}
\end{equation}
where
\begin{equation}
\bar{\partial}_{J}u:=\frac{1}{2}(du+J\circ du\circ j) \label{AntiholomorphicTerm}
\end{equation}
is the antiholomorphic part of $du$. 
The identity (\ref{energyequality}) shows that 
\begin{equation}
\frac{1}{2}\int_{\Sigma} |du|^2_{g_J} dA_{\Sigma} \ge \int_{\Sigma} u^{*}\omega\,,
\end{equation}
with equality holding if and only if  $\bar{\partial}_{J}u=0$,
or equivalently, iff (\ref{JHolomorphicCondition}) holds.
Because $J$-holomorphic maps
minimize energy, they are harmonic maps. For general smooth maps 
$u: \Sigma \rightarrow \calM$, the energy $E(u)$ depends on the metric $g_{J}$ on $\calM$. 
However, the energy identity shows that, for $J$-holomorphic curves in symplectic manifolds, 
$E(u)$ equals the minimal area $V(u)$ and that this quantity
is a topological invariant that depends only on the homology class of the curve modulo its 
boundary.\cite{McS04}

Our inspiration to use $J$-holomorphic maps for
investigating the quantum uncertainty principle comes from Oh's lecture notes,\cite{Oh95}
wherein he suggests that Gromov's non-squeezing theorem is a
classical analogue of the quantum uncertainty principle. 
With an eye to the similarities of classical and quantum mechanics,
we focus on the the Robertson-Schr\"{o}dinger uncertainty relation, which states that
\begin{equation}
(\Delta \hat{A})_{\psi}^{2}\,(\Delta \hat{B})_{\psi}^{2}
 - (C(\hat{A},\hat{B})_{\psi})^{2} \geq
\left(\frac{1}{2i} \langle [\hat{A},\hat{B}] \rangle_{\psi} \right)^2,\label{Shankarprinciple}
\end{equation}
where $\psi \in \calH$ and $\Delta \hat{A}$, $\Delta \hat{B}$, and $C(\hat{A},\hat{B})$ are the 
uncertainty and covariance functions for self-adjoint linear operators $\hat{A}$ and $\hat{B}$ on $\calH$.
By using $\hat{A}$ and $\hat{B}$ to define a family of maps $\mathcal{F}$ from a Riemann surface 
$\Sigma$ into a finite-dimensional phase space $P(\calH)$, we find that
the uncertainty relation (\ref{Shankarprinciple})
is equivalent to the differential version of the energy identity, 
\begin{equation}
\frac{1}{2} |du|_{g}^{2} dA_{\Sigma}\ge  u^{*} \Omega,\label{eidentity}
\end{equation}
where $g$ and $\Omega$ are, respectively, the Riemannian metric and the symplectic form on $P(\calH)$,
and $u$ is a map in $\mathcal{F}$. When $u$ is holomorphic, the inequality is saturated and the 
off-diagonal covariance term $C(\hat{A},\hat{B})$ vanishes. For a compact Riemann surface
where such a map $u: \Sigma \rightarrow P(\calH)$ can be globally defined, 
the integral form of the uncertainty inequality can be interpreted as a comparison of 
 the area of $\Sigma$ as measured in the pull-back metric $u^{*}(g)$, 
 to the symplectic area of $\Sigma$, as measured by
pulling back the symplectic form on $P(\calH)$. In this case, if the map is holomorphic, the
 uncertainty product integral over $\Sigma$ is a topological invariant within the homology class 
 of the curve.

The remainder of the paper is organized as follows.
Section 2 establishes our notation and viewpoint by reviewing the theory of 
geometric quantum mechanics, and casts the Robertson-Schr\"{o}dinger inequality 
in terms of the symplectic form and Riemannian metric on $P(\calH)$.
Section 3 proves our claim that (\ref{Shankarprinciple}) is an example of (\ref{eidentity}) by 
defining the map family $\mathcal{F}$, showing 
that the Fubini-Study metric tensor pulls back by a map in $\mathcal{F}$ to the quantum covariance tensor, 
and computing the energy differential and symplectic form in the pull-back metric.

\section{Quantum mechanics as a Hamiltonian dynamical system}

A Hamiltonian dynamical system $(\mathcal{M},\omega,X_{E})$ consists of a phase space, $\calM$, which is a
smooth manifold equipped with a symplectic form, $\omega$,
and a preferred hamiltonian vector field, $X_{E}: \calM \rightarrow T(\calM)$. 
A smooth two-form on $\calM$ is symplectic if it is closed and nondegenerate.
A vector field $X$ on $\calM$ is hamiltonian if $\iota_{X}
\omega$ is exact, where
$\iota_{X} \omega (\cdot):=\omega(X,\cdot)$, or equivalently, if there is a
$C^{1}$ function $F: \calM \rightarrow \mathbb{R}$ such that
\begin{equation}
\iota_{X}\omega=dF.\label{Hamvectorfield}
\end{equation}
When (\ref{Hamvectorfield}) holds, we write $X=X_{F}$. Each point of the symplectic
manifold $\mathcal{M}$ corresponds to a state of the physical
system, and the time evolution of the system is given by the flow
along the preferred vector field $X_{E}$. With initial condition $z_{m}(0)=m
\in \calM$, the trajectory $z_{m}(t)$ is uniquely determined by
Hamilton's equations,
\begin{equation}
\frac{d z}{dt} = (X_{E})_{z(t)}.\label{Hamiltonseqs}
\end{equation}
The observables, or measurable quantities, in Hamiltonian
mechanics are real-valued differentiable functions on $\calM$.
For two observables $G,F:\calM \rightarrow \mathbb{R}$, the Poisson bracket $\{\,,\,\}$
is defined as 
 \begin{equation}
\{G,F\}:=\omega(X_{G},X_{F})=dG(X_{F}). \label{Poisson_bracket}
\end{equation}
 The time evolution of an observable $G$
is given by $ dG/dt=\{G,E\}$, where $E$ is the energy function corresponding to the preferred vector
field $X_{E}$ with $\iota_{X_{E}}\omega=dE$.

Recall that the traditional algebraic formulation of quantum mechanics describes a quantum system 
in terms of a complex separable Hilbert space $\calH$ with
Hermitian inner product $\langle \,\,, \,\, \rangle$, and that the
observables of the system are represented by self-adjoint linear
operators on $\calH$. A special role is played by the Hamiltonian operator, $\hat{H}$,
whose eigenvalues are the energies of the system: $\hat{H}
\psi_{\lambda}=E_{\lambda} \psi_{\lambda}$. In the Heisenberg
picture, linear operators on $\calH$ time evolve according to
$d \hat{A}/dt=(i/\hbar)[\hat{H},\hat{A}]$,
where the commutator of $\hat{A}$ and $\hat{B}$
is $[\hat{A},\hat{B}]=\hat{A}\hat{B}-\hat{B}\hat{A}$.
In the Schr\"{o}dinger picture, the operators are stationary, but
the elements of $\calH$ time-evolve according to Schr\"{o}dinger's
equation,
\begin{equation}
\frac{d\psi (t)}{dt}=-\frac{i}{\hbar}\hat{H}\psi(t).\label{Scheqn}
\end{equation}

Geometric quantum mechanics reformulates the algebraic quantum theory in geometric terms
by regarding the linear space $\calH$ as a Hermitian manifold, with its tangent bundle $T(\calH)$
 identified with $\calH \times
\calH$. For each $\psi \in \calH$, the canonical isomorphism of
$\calH$ onto $T_{\psi}(\calH)$, $v \mapsto v_{\psi}$, is given by
$v_{\psi} := \alpha ^{\prime}(0)$, where $\alpha : \mathbb{R}
\rightarrow \calH$ is defined by $\alpha (t)=\psi+tv$. 
The pair $(\psi,v) \in \calH \times \calH$ corresponds to 
$v_{\psi} \in T_{\psi}(\calH)$.
The inner product of two vectors $v_{\psi}$ and $w_{\psi}$ in $T_{\psi}(\mathcal{H})$ is
defined by $\langle v_{\psi},w_{\psi} \rangle _{T(\calH)}:=\langle v,w \rangle$, 
where $\langle\cdot,\cdot \rangle$ is the Hermitian inner product on $\calH$.
The natural symplectic form $\Omega_{\mathcal{H}}$ on $\calH$ is given by the imaginary part of this inner product. For $v_{\psi}, w_{\psi} \in T_{\psi}(\mathcal{H})$,
\begin{equation}
(\Omega_{\mathcal{H}})(v_{\psi},w_{\psi}):=2 \hbar\,{\rm Im}
\langle v,w \rangle.\label{OmHdefinition}
\end{equation}
For a linear operator $\hat{A}$ on $\mathcal{H}$, the expectation
value function $A: \mathcal{H} \rightarrow \mathbb{R}$ 
\begin{equation}
A(\psi):=\langle \hat{A} \rangle _{\psi}:=\frac{\langle \psi,\hat{A}\psi \rangle}{\langle \psi,\psi
\rangle}\,\label{fun}
\end{equation}
is used to define the hamiltonian vector
field $X_{A}$ on $\mathcal{H}$ given by $\iota_{X_{A}}\Omega_{\mathcal{H}}=dA$, where  
\begin{equation}
(X_{A})_{\psi}:=\frac{-i}{\hbar}\hat{A}\psi. \label{XAdef}
\end{equation}
Thus, the symplectic form $\Omega_{\calH}$ acting on two hamiltonian vector fields $X_A, X_B \in T(\calH)$
is proportional to the expectation value of the
commutator of the self-adjoint operators $\hat{A}$ and $\hat{B}$ on $\calH$, 
\begin{equation}
(\Omega_{\calH})_{\psi}(X_A,X_B)
=-\frac{i}{\hbar}\left \langle [\hat{A},\hat{B}] \right\rangle _{\psi}.\label{commsymp}
\end{equation} 
Quantum dynamics can be described\cite{MR} 
in terms of the flow along the preferred hamiltonian vector field $X_{H}$ on $\mathcal{H}$, where 
$\hat{H}$ is the Hamiltonian operator for the system. 
Schr\"{o}dinger's equation (\ref{Scheqn}) on $\calH$ takes
the form of Hamilton's equations (\ref{Hamiltonseqs}),
\begin{equation}
\frac{d\psi}{dt}=(X_{H})_{\psi(t)}\,.
\end{equation}

Nevertheless, it is the complex projective space $P(\calH$) rather than $\mathcal{H}$
that is generally regarded as the true  phase space of geometric quantum mechanics, for the
following reason. As is clear from the definition (\ref{fun}), 
the expectation value of $\hat{A}$ on $\psi$ is equal to the expectation value
of $\hat{A}$ on $c\psi$, for any nonzero $c \in \mathbb{C}$. Thus, a
pure quantum state must be regarded as a complex line through the origin
in $\calH$, or an equivalence class [span$\{\psi\}] \in
P(\calH):=\calH\diagup{\sim}$, where $\psi \sim \psi'$ if and only if
$\psi'=c\psi$ for some nonzero $c \in \mathbb{C}$. 
Hereafter, we use the abbreviated notation $[\psi]$ to denote the equivalence class [span$\{\psi\}]$.
State vectors $\psi \in \calH-\{0\}$ and  physical states $[\psi] \in P(\calH)$ 
are related by means of the principal bundle
$\mathbb{C}^{\ast}\hookrightarrow (\calH -\{0\}) \xrightarrow{\pi}
P(\calH)$, where $\mathbb{C}^{\ast}$ is the multiplicative group of nonzero complex numbers and the projection map $\pi$ is defined by $\pi(\psi)=[\psi].$
The Hermitean inner product $\langle \cdot , \cdot \rangle$ on $\calH$ provides a natural
principal connection on the bundle. 
For each $\psi \in \calH -\{0\}$, the {\em horizontal} {\em subspace} of 
$T_{\psi}(\calH -\{0\})$ consists of
all $v_{\psi}=(\psi,v)$ such that $\langle \psi,v \rangle = 0$.
The tangent map $\pi_*$ is an isomorphism
from the horizontal subspace of $T_{\psi}(\calH -\{0\})$ onto
$T_{[\psi]}(P(\calH))$.
 
Pushing forward the inner product $\langle \cdot , \cdot \rangle$ on $\calH$ by  
$\pi_{*}$ induces an Hermitian inner product 
$\langle \langle \cdot , \cdot \rangle \rangle$ on $P(\calH)$, as follows.\cite{Arnold}
Any $v_{\psi} \in T_{\psi}(\calH-\{0\})$ can be identified with
$v=c\psi+\delta \in \calH$ where $c=\langle v,\psi
\rangle/\langle \psi,\psi \rangle$ and
$\delta=v-c\psi.$ Thus, $\pi_{\ast}(v_{\psi})=\pi_{\ast}(\delta_{\psi})$
and $\langle \delta,\psi \rangle =0$. To define
$\langle \langle \cdot , \cdot \rangle \rangle$ on $P(\calH)$, 
let $p =[\psi] \in P(\calH)$ and $V,W \in T_{p}(P(\calH))$. 
Choose any $\psi \in \pi^{-1}(p)$ and $v_{\psi},w_{\psi} \in T_{\psi}(\calH)$ satisfying
$\pi_{*}v_{\psi}=V$, $\pi_{*}w_{\psi}=W$.
The scaling transformation $\psi \rightarrow \psi/|\psi|$ is required
 to ensure that the inner product of $V$
and $W$ does not depend on which element of $\pi^{-1}(p)$ is chosen.
 The Hermitian inner product of $V$ and $W$ on $T(P(\calH))$ is computed by taking the inner product
of the horizontal components of $v$ and $w$,
\begin{equation}
\langle \langle V,W \rangle \rangle :=
\left \langle
v-\frac{\langle v, \psi \rangle}{\langle \psi,\psi
\rangle}\psi,w - \frac{\langle w, \psi \rangle}{\langle
\psi,\psi \rangle}\psi  \right \rangle  \langle \psi,\psi
\rangle^{-1}.
 \label{FSmetric}
\end{equation}

Geometric quantum mechanics uses the Hermitian inner product (\ref{FSmetric}) to describe quantum dynamics
in terms of the symplectic geometry of 
$P(\calH)$, regarded as the phase space for a Hamiltonian
system.\cite{AS, Boya, Cirelli90, Kibble, Schilling}  The symplectic
form on $P(\calH)$ is given by
\begin{equation}
\Omega (V,W):=2 \hbar\,{\rm Im} \langle \langle V,W \rangle \rangle.\label{qsymform}
\end{equation}
Since the expectation value $A(\psi)$ in (\ref{fun}) does not depend on the choice of element
of span\{$\psi\}$, the quantum observable $a:P(\calH) \rightarrow \mathbb{R}$
is well-defined as
\begin{equation}
a([\psi]):=A(\psi)=\frac{\langle \psi,\hat{A}\psi \rangle}{\langle \psi,\psi
\rangle}\,, \label{observable}
\end{equation}
with corresponding hamiltonian vector field $X_{a}$ on $P(\mathcal{H})$
 defined by $\iota_{X_{a}}\Omega=da$.
The tangent map $\pi_{\ast}: T(\calH-\{0\}) \rightarrow T(P(\calH))$ gives
\begin{equation}
X_a=\pi_{*}(X_A) \label{horizvectfield}
\end{equation}
with $X_A$ as defined in (\ref{XAdef}). The
dynamics on the phase space $P(\mathcal{H})$ follows the flow of the preferred hamiltonian vector
field $X_{h}$, which is the push-forward 
 $X_{h}=\pi_{*} (X_H)$, where $\hat{H}$ is the Hamiltonian operator on $\calH$.
Thus, similarly to classical mechanics, observables in geometric quantum mechanics are
differentiable real-valued functions on a nonlinear symplectic manifold, 
while the dynamics of the system's states is determined by Hamilton's
equation. Quantum mechanics is a special Hamiltonian dynamical system, since  
the real part of (\ref{FSmetric}) serves as a Riemannian metric on $P(\calH)$,
\begin{equation}
g(V,W):=2 \hbar\,{\rm Re} \langle \langle V,W \rangle \rangle.
\label{metric}
\end{equation}
If $\calH$ is regarded as $\mathbb{C}^{n+1}$,
then $g$ is a constant multiple of the Fubini-Study metric on $P(\mathbb{C}^{n+1})=\mathbb{C}$P$^n$.

Both the symplectic and Riemannian parts of the Hermitian structure
(\ref{FSmetric}) on $P(\calH)$ are key to a geometric understanding of the 
uncertainty relation (\ref{Shankarprinciple}). Let $\hat{A}$ and $\hat{B}$ be 
self-adjoint linear operators on $\calH$ and let $\psi \in S(\calH)$, where 
$S(\calH)=\{\psi \in \calH : |\psi|^{2}=1\}$. 
The uncertainty or dispersion in the values of $A$ is given by the function 
$\Delta \hat{A}: S(\calH) \rightarrow \mathbb{R}$, defined as
\begin{equation}
(\Delta \hat{A})_{\psi}:=[\langle \psi, (\hat{A}-\langle \hat{A}
\rangle_{\psi})^{2} \psi \rangle ]^{1/2},\label{dispersion}
 \end{equation}
 where $\langle
\hat{A} \rangle _{\psi}$ is the expectation value function (\ref{fun}).
The covariance or correlation function $C(\hat{A},\hat{B}): S(\calH) \rightarrow \mathbb{R}$ is defined as
\begin{equation}
C(\hat{A},\hat{B})_{\psi} :=
\frac{1}{2}\left\langle
\psi,\left[(\hat{A}-\langle\hat{A}\rangle
_{\psi})(\hat{B}-\langle\hat{B}\rangle
_{\psi})+(\hat{B}-\langle\hat{B}\rangle
_{\psi})(\hat{A}-\langle\hat{A}\rangle _{\psi})\right]\psi
\right\rangle.
\label{covariance}
\end{equation}
 A geometric interpretation of (\ref{Shankarprinciple})
  comes from recognizing that $(\hat{A}-\langle  \hat{A}  \rangle _{\psi}) \psi$ 
is the horizontal component of $\hat{A}\psi$ by the Hermitian connection on the canonical 
$U(1)$ bundle over $P(\calH)$. 
\begin{lemma}\cite{Anan90} Let $\hat{A}$
be a linear operator on $\calH$ and let $\psi \in S(\calH)$.\\
1) The component of the vector
$\hat{A}\psi \in \calH$ that is Hermitian orthogonal to $\psi$ is
$(\hat{A}-\langle  \hat{A}  \rangle _{\psi}) \psi$.\\
2) If $\hat{A}$ is self-adjoint, then $\hat{A}\psi$
decomposes as
$\hat{A}\psi=\langle \hat{A} \rangle_{\psi} \psi + (\Delta
\hat{A})_{\psi} \chi$,
 where $\langle \chi, \psi
\rangle =0$ and $\langle \chi, \chi \rangle =1$.
\end{lemma}
\begin{cor}
A hamiltonian vector field $X_A$ on $S(\calH)$ decomposes into vertical and horizontal components,
 $X_A=X_A^{vert} + X_A^{horiz}$, where
\begin{eqnarray}
& &(X_A^{vert})_{\psi}=\frac{-i}{\hbar}\langle \hat{A} \rangle_{\psi} \psi, \nonumber \\
& & (X_A^{horiz})_{\psi}=\frac{-i}{\hbar}(\hat{A}-\langle \hat{A} \rangle _{\psi}\hat{I}) \psi,
 \label{horizvectorfield}
\end{eqnarray}
and $\hat{I}$ is the identity operator on $\calH$.
The inner product (\ref{FSmetric}) of hamiltonian vector fields $X_a$ and $X_b$ on $P(\calH)$ is given by
\begin{equation}
\langle \langle X_a,X_b \rangle \rangle = \langle X_A^{horiz}, X_B^{horiz} \rangle. \label{inner_products}
\end{equation}
\end{cor}

The uncertainty relation (\ref{Shankarprinciple}) is proved in standard texts\cite{Shankar} by using 
the Cauchy-Schwartz inequality. 
The relation can be cast in a geometric form by writing it in terms of
the real and imaginary parts of the Hermitian structure on $P(\calH)$.\cite{AS, BH, Cirelli90,
Schilling}.
\begin{cor}
Let $\hat{A}$ and $\hat{B}$ be self-adjoint linear operators on $\calH$ and let $\psi \in S(\calH)$.
The Robertson-Schr\"{o}dinger uncertainty relation (\ref{Shankarprinciple}) 
has the following equivalent geometric expression
in terms of the symplectic form $\Omega$ and the Riemannian metric $g$ on $P(\calH)$,
\begin{equation}
g_{[\psi]}(X_a,X_a) g_{[\psi]}(X_b,X_b)-(g_{[\psi]}(X_a,X_b))^2
 \ge \left( 
 \Omega_{[\psi]}(X_a,X_b) \right)^2,\label{geounprinciple}
 \end{equation}
\end{cor}
\pf
 By the self-adjoint property of $\hat{A}$, 
\[(\Delta \hat{A})_{\psi}^{2}= \langle (\hat{A}-\langle \hat{A}
\rangle _{\psi}) \psi, (\hat{A}-\langle  \hat{A}  \rangle _{\psi}) \psi \rangle.\]
Using (\ref{horizvectorfield}) and (\ref{inner_products}),
\begin{equation}
(\Delta \hat{A})_{\psi}^{2}=\hbar^2 \langle X_A^{horiz},X_A^{horiz} \rangle _{\psi}
=\hbar^2 \langle \langle X_a,X_a \rangle \rangle _{[\psi]},\label{delAsq}
\end{equation}
which is clearly real, so that by the definition (\ref{metric}) of $g$
\begin{equation}
g_{[\psi]}(X_a,X_a)= \frac{2}{\hbar} (\Delta \hat{A})_{\psi}^{2}. \label{uncertainty_metric}
\end{equation}
Similarly,
\begin{eqnarray}
& & g_{[\psi]}(X_b,X_b) =\frac{2}{\hbar}(\Delta \hat{B})_{\psi}^{2} ,\\
& &g_{[\psi]}(X_a,X_b)=2\hbar\,{\rm Re} \langle \langle X_a,X_b \rangle \rangle
=\frac{2}{\hbar} C(\hat{A},\hat{B})_{\psi}.
\label{covariance_metric}
\end{eqnarray}
The commutator term on the right hand side of (\ref{Shankarprinciple}) satisfies
\[\langle \psi,[\hat{A},\hat{B}]\psi \rangle = \langle\psi,[(\hat{A}-\langle\hat{A}\rangle),(\hat{B}-\langle\hat{B}\rangle)]\psi\rangle,\]
so that, by using  the relation (\ref{commsymp}) between the commutator and the symplectic form on 
$\calH$, as well as (\ref{horizvectorfield}), (\ref{inner_products}), and the definition (\ref{qsymform})
of $\Omega$,
\begin{equation}
\Omega_{[\psi]}(X_a,X_b) = -\frac{i}{\hbar} \langle [\hat{A},\hat{B}] \rangle_{\psi}\,.
 \end{equation}
\endpf

\section{The uncertainty principle and the energy identity}

This section states and proves the claim that the Robertson-Schr\"{o}dinger uncertainty relation 
is an example of the differential version of the energy identity for $J$-holomorphic maps from a Riemann 
surface into the quantum state space $P(\calH)$.
Beginning with the result from the previous section that the uncertainty relation can be expressed in 
terms of the symplectic form $\Omega$ and Riemannian metric $g$ on $P(\calH)$, observe that the 
inequality can be interpreted as a minimal area condition on parallelograms formed by 
non-commuting vector fields at a point in $P(\calH)$.\cite{Sanborn}
\begin{prop} 
Let $\hat{A}$ and $\hat{B}$ be self-adjoint linear operators on $\calH$ and let $\psi \in S(\calH)$.
Define the covariance tensor $M(\hat{A},\hat{B})$ corresponding to the measurement of $\hat{A}$ and $\hat{B}$ as
\begin{equation}
M(\hat{A},\hat{B})_{\psi}:=\frac{2}{\hbar}\left( \begin{array}{cc}
(\Delta \hat{A})_{\psi}^{2} & C(\hat{A},\hat{B})_{\psi} \\
C(\hat{B},\hat{A})_{\psi} &  (\Delta \hat{B})_{\psi}^{2}
 \end{array} \right).\label{covariance_matrix}
 \end{equation}
The determinant of the matrix $M(\hat{A},\hat{B})_{\psi}$ is bounded below by the square of the 
symplectic area of the parallelogram formed by the projected vectors $(X_a)_{[\psi]}$
and $(X_b)_{[\psi]}$ in $T_{[\psi]}(P(\calH))$;
\begin{equation}
{\rm det}\,M(\hat{A},\hat{B})_{\psi} \ge \left(\Omega_{[\psi]}(X_a,X_b) \right)^2.\label{detprinciple}
 \end{equation}
\end{prop}
\pf
 By the proof of Corollary 3, the 
elements of $M(\hat{A},\hat{B})_{\psi}$ can be
 expressed in terms of the metric $g$ on $P(\calH)$, so that
\begin{equation}
M(\hat{A},\hat{B})_{\psi}=\left( \begin{array}{cc}
g_{[\psi]}(X_a,X_a) & g_{[\psi]}(X_a,X_b) \\
g_{[\psi]}(X_b,X_a) &  g_{[\psi]}(X_b,X_b)
 \end{array} \right).
 \end{equation}
 The result then follows by observing that $\Omega_{[\psi]}(X,Y)$ measures the differential
 area element formed by vectors $X,Y \in T_{[\psi]} (P(\calH))$.
 \endpf
 
 The proof shows that $M(\hat{A},\hat{B})$ has the form of a Riemannian metric tensor on a two-dimensional
manifold, with components determined by the metric $g$ on $P(\calH)$. Accordingly,
$M(\hat{A},\hat{B})$ acts as the metric on a Riemann surface $\Sigma$, given as the 
pull-back of $g$ by a map $u: \Sigma \rightarrow P(\calH)$ using the vector fields $X_a$ and $X_b$.
To characterize maps that are suitable for investigating the uncertainty relation by using 
$J$-holomorphic curves, it is sufficient to specify a condition on the map differential, since
the terms of the energy identity and the uncertainty inequality depend only on operations 
involving tangent vectors.  
Our analysis assumes the standard spaces $\calH=\mathbb{C}^{n+1}$, $S(\calH)={\rm S}^{2n+1}$, and $P(\calH)=\mathbb{C}$P$^n$. In local coordinates $(y^{1},\cdots,y^{2n+2})$ on 
$\calH \simeq \mathbb{R}^{2n+2}$, the standard complex structure is given by
$J(\partial/\partial y^i)=\partial /\partial y^{i+1}$ ($i$ odd)
and $J(\partial/\partial y^i)=-\partial /\partial y^{i+1}$ ($i$ even) so that
$J$ is the $(2n+2) \times (2n+2)$ 
block diagonal matrix with each block equal to the $2 \times 2$ matrix $j_0$, where
\begin{equation}
j_0:= \left(
\begin{array}{cc}
0 & -1 \\
1 & 0
 \end{array} \right).
\end{equation}
We assume without loss of generality that $\Sigma$ is an open subset of $\mathbb{C}$ 
parameterized by $x=s+it$.
\begin{definition}
Let $\hat{A}$ and $\hat{B}$ be self-adjoint linear operators on $\calH$ and let 
 $(s,t)$ be holomorphic coordinates on $\Sigma$. Define the family of maps, 
\begin{eqnarray}
\mathcal{F}&:=&\{u: \Sigma \rightarrow P(\calH) \,|\, \exists f:\Sigma \rightarrow S(\calH)
\,\,{\rm such}\,\,{\rm that}\,\, u=\pi \circ f \nonumber \\
& &{\rm and}\,\,(df)_x(\partial_{s}) =(X_{A}^{horiz})_{f(x)},\,
(df)_{x}(\partial_{t}) =(X_{B}^{horiz})_{f(x)}\},
 \label{MapFamily}
\end{eqnarray}
where $(\partial_{s},\partial_{t})$ abbreviates 
$(\frac{\partial}{\partial s},\frac{\partial}{\partial t})$, and
$\pi$ is the  projection from $\calH$ onto $ P(\calH)$. 
\end{definition}

\begin{lemma}
Let $\hat{A}$ and $\hat{B}$ be self-adjoint linear operators on $\calH$.
If $u: \Sigma \rightarrow P(\calH)$ is a map in the family $\mathcal{F}$, 
then the Fubini-Study metric tensor $g$
on $P(\calH)$ pulls back by $u$ to the covariance tensor defined in (\ref{covariance_matrix}). 
That is, the induced metric tensor $h=u^{*}(g)$ on $\Sigma$
is given by $(h_{k \ell})=M(\hat{A},\hat{B})$. 
\end{lemma}
\pf
Let $u: \Sigma \rightarrow P(\calH)$ be a map in the family $\mathcal{F}$.
Then $u=\pi \circ f$ for some $f: \Sigma \rightarrow S(\calH)$ with the property  
(\ref{MapFamily}). In a linear orthonormal basis $\{\phi_{\alpha}\}_{\alpha=1}^{n+1}$, let
$\tilde{g}$ be the Riemannian metric defined by
$\tilde{g}(\phi_{\alpha},\phi_{\beta}):=2 \hbar \delta_{\alpha \beta}$ on 
$\mathbb{C}^{n+1}$. The first step is to show that the pull-back of the Fubini-Study metric $g$ 
on $P(\calH)$ by the map $u$ 
is equal to the pull-back of the metric $\tilde{g}$ on 
$\calH$ by the map $f$. 
Let $\xi, \zeta \in T_x(\Sigma)$. 
By the definition (\ref{metric}) of the metric $g$ on $P(\calH)$,
\begin{eqnarray}
u^{*}(g)(\xi, \zeta)&=&g(u_{*}\xi,u_{*} \zeta) \nonumber \\
& =& 2 \hbar {\rm Re} \langle \langle u_{*} \xi,u_{*} \zeta \rangle \rangle. \nonumber 
\end{eqnarray}
The differential $df_x$ maps vectors in $T_x(\Sigma)$ into the horizontal subspace of 
$T_{f(x)}(S(\calH))$. Hence, by the definition (\ref{FSmetric}) of the inner product 
$\langle \langle \cdot,\cdot \rangle \rangle$ on $P(\calH)$,
\begin{eqnarray}
2 \hbar {\rm Re} \langle \langle u_{*} \xi,u_{*} \zeta \rangle \rangle
&= & 2 \hbar {\rm Re} \langle f_{*} \xi,f_{*} \zeta\rangle \nonumber \\
&=& f^{*}(\tilde{g})(\xi, \zeta) \nonumber
\end{eqnarray}
Therefore,
\begin{equation}
u^{*}(g)=f^{*}(\tilde{g}).\label{equalpullbacks}
\end{equation}
The second step is to show that $(f^{*}(\tilde{g}))_{x}=M(\hat{A},\hat{B})_{f(x)}$.
In coordinates $(x^1,x^2)=(s,t)$ on $\Sigma$, the pull-back metric $h=u^{*}(g)$ on $\Sigma$ is
\[h=\sum_{k, \ell}\, h_{k \ell}\, dx^{k}dx^{\ell}.\]
By (\ref{equalpullbacks}),
\begin{eqnarray}
h_{k \ell}
&=&(f^{*}\tilde{g})\left(\frac{\partial}{\partial x^{k}},
\frac{\partial}{\partial x^{\ell}}\right) 
=\tilde{g}\left(\frac{\partial f}{\partial x^{k}},
\frac{\partial f}{\partial x^{\ell}}\right) \nonumber\\
&=&\tilde{g}\left(\sum_{\alpha} \frac{\partial f^{\alpha}}{\partial x^{k}}\phi_{\alpha},
\sum_{\beta}\frac{\partial f^{\beta}}{\partial x^{\ell}}\phi_{\beta}\right)
=2 \hbar {\rm Re}\sum_{\alpha}\left(\overline{\frac{\partial f^{\alpha}}{\partial x^{k}}}\right)
\frac{\partial f^{\alpha}}{\partial x^{\ell}}\,. \nonumber
\end{eqnarray}
By (\ref{MapFamily}), $\frac{\partial f}{\partial s}=X_{A}^{horiz}$ and
$\frac{\partial f}{\partial t}=X_{B}^{horiz}$. Thus, 
\[h_{11}=2 \hbar \left\lvert\frac{\partial f}{\partial s}  \right\rvert ^2
=2 \hbar \,\langle X_A^{horiz}, X_A^{horiz} \rangle.\]
By (\ref{delAsq}),
$(h_{11})_x=\frac{2}{\hbar} (\Delta \hat{A})_{f(x)}^{2}$.
Similarly,
$(h_{22})_x=\frac{2}{\hbar}(\Delta \hat{B})_{f(x)}^{2}$ and 
$(h_{12})_x=(h_{21})_x=2 \hbar{\rm Re}\langle X_{A}^{horiz},X_{B}^{horiz} \rangle_{f(x)}=
\frac{2}{\hbar}C(\hat{A},\hat{B})_{f(x)}$. By comparing to the definition 
(\ref{covariance_matrix}) of the covariance matrix, the result is obtained.
\endpf

\begin{theorem} \cite{Sanborn}
Let $\hat{A}$ and $\hat{B}$ be self-adjoint linear operators on $\calH$.
Let $u$ be a map in the family $\mathcal{F}$ with $u=\pi \circ f$ and $f:\Sigma \rightarrow S(\calH)$
with $f(x)=\psi$.
Then, in the pull-back metric $h=u^{*}(g)$ on $\Sigma$,
\begin{equation}
\frac{1}{2} (|du|^2 dA_{\Sigma})_x= \frac{2}{\hbar}\sqrt{(\Delta \hat{A})_{\psi}^{2}\,(\Delta
\hat{B})_{\psi}^{2}-(C(\hat{A},\hat{B})_{\psi})^{2} }\,  ds \wedge dt,
\end{equation}
 \begin{equation}
 (u^{*}(\Omega))_x=\frac{1}{2i}\langle \psi, [\hat{A},\hat{B}]\psi \rangle\, ds \wedge dt.\label{RHS}
 \end{equation}
Thus, the Robertson-Schr\"{o}dinger uncertainty relation (\ref{Shankarprinciple})  
is equivalent to the differential version of the energy identity (\ref{eidentity}).
\end{theorem}
\pf
The first step is to show that choosing the pull-back metric $h=u^{*}(g)$ on $\Sigma$ to compute the
energy density gives $e(u)=\frac{1}{2} \mid du \mid^2=1.$
In coordinates, $(x^{1},x^{2})=(s,t)$ on $\Sigma$, writing 
$h=\sum h_{k \ell}dx^{k}dx^{\ell}$ and
$(h^{k \ell})=(h_{k \ell})^{-1}$,
\begin{equation}
|du|^{2}:=\sum_{k, \ell,\alpha,\beta} g_{\alpha \beta}
(u(x))h^{k \ell}(x) \frac{\partial u^{\alpha}}{\partial x^{k}}
\frac{\partial u^{\beta}}{\partial x^{\ell}}\,.\label{dusquared1}
\end{equation}
Equivalently, again using the fact that $f_{*}$ maps into the 
horizontal subspace of $T_{\psi}(S(\calH))$, 
\begin{equation}
|du|^{2}=\sum_{k, \ell,\alpha,\beta} \tilde{g}_{\alpha \beta}
(f(x))h^{k \ell}(x) \frac{\partial f^{\alpha}}{\partial x^{k}}
\frac{\partial f^{\beta}}{\partial x^{\ell}}\,,\label{du2}
\end{equation}
where $\tilde{g}_{\alpha \beta}=2 \hbar\delta_{\alpha \beta}$.
Inverting $(h_{k \ell})$ and using (\ref{du2}) with $s=x^{1}$ and $t=x^{2}$, we have
\begin{equation}
|du|^{2}=2 \hbar
 \left[ h^{11}\left\lvert\frac{\partial f}{\partial s}  \right\rvert ^2
+h^{12}{\rm Re}\sum_{\alpha}\left(\overline{\frac{\partial f^{\alpha}}{\partial s}}\right)
 \frac{\partial f^{\alpha}}{\partial t}
+h^{21}{\rm Re}\sum_{\alpha}\left(\overline{\frac{\partial f^{\alpha}}{\partial t}}\right) 
\frac{\partial f^{\alpha}}{\partial s}
+h^{22}\left\lvert\frac{\partial f}{\partial t} \right\rvert ^2 \right]
\end{equation}
\begin{eqnarray}
 |du|^{2}&=&\frac{1}{{\rm det}(h_{k \ell})} (2h_{11}^{2} h_{22}^{2}-2h_{12}^{2}) \nonumber \\
 &=&2.\label{dusquared2}
\end{eqnarray}
Thus, the left hand side of the differential energy identity (\ref{eidentity}) is
\begin{eqnarray}
\frac{1}{2}|du|^2dA_{\Sigma}&=&dA_{\Sigma}=\sqrt{{\rm det}(h_{ij})_{x}}\,ds \wedge dt\nonumber\\
&=&\frac{2}{\hbar}\sqrt{(\Delta \hat{A})_{\psi}^{2}\,(\Delta
\hat{B})_{\psi}^{2}-(C(\hat{A},\hat{B})_{\psi})^{2} }\,ds \wedge dt\,.
\end{eqnarray}

To prove (\ref{RHS}), choose symplectic coordinates $(y_{1},\cdots,y_{2n+2})$ in a neighborhood of 
$\psi =f(x)\in \calH$.
 Define the horizontal vectors $v,w \in T_{\psi}(S(\calH))$ as  
\begin{eqnarray}
& &v:=(X_A^{horiz})_{\psi}=-\frac{i}{\hbar}(\hat{A}-\langle \hat{A} \rangle_{\psi} )\psi, \nonumber\\
& &w:=(X_B^{horiz})_{\psi}=-\frac{i}{\hbar}(\hat{B}-\langle \hat{B} \rangle _{\psi})\psi.
\label{vectors}
\end{eqnarray}
Then
\begin{eqnarray}
f(s,t)&=& (y_{1}(s,t),\cdots,y_{2n}(s,t))^{T}\nonumber\\
f^{*}(dy_{\alpha})&=&(v_{\alpha}ds+w_{\alpha}dt),\,\,\,\,\,\alpha=1,\cdots,2n.
\end{eqnarray}
$u^{*}\Omega =f^{*} \Omega_{\calH}$, so that  
\begin{eqnarray}
(u^{*}\Omega)_x&=&f^{*} \left(\sum_{\alpha\,\, {\rm odd}} dy_{\alpha}\wedge dy_{\alpha+1} \right) 
=\sum_{\alpha\,\, {\rm odd}}f^{*}(dy_{\alpha})\wedge f^{*}(dy_{\alpha+1})\nonumber
\\
&=&\sum_{\alpha\,\, {\rm odd}}(v_{\alpha}ds+w_{\alpha}dt) \wedge (v_{\alpha+1}ds+w_{\alpha+1}dt)
\nonumber\\
&=&\sum_{\alpha\,\, {\rm odd}}(v_{\alpha}
w_{\alpha+1}-v_{\alpha+1}w_{\alpha})ds\wedge dt\nonumber\\
&=&\sum_{\alpha\,\, {\rm odd}} {\rm det} \left( \begin{array}{cc}
v_{\alpha} & w_{\alpha} \\
 v_{\alpha+1} & w_{\alpha+1}
 \end{array} \right)ds \wedge dt\\
 &=&\langle v,J w \rangle\,ds \wedge dt = \Omega_{\calH}(v,w)\, ds \wedge dt\\
 &=& \frac{1}{2i}\langle \psi, [\hat{A},\hat{B}]\psi \rangle\, ds \wedge dt,
\end{eqnarray}
where we have used the relation (\ref{commsymp}).
\endpf
 
 \begin{cor}
If $u \in \cal{F}$ is $J$-holomorphic, then $X_b=J X_a$.
\end{cor}
\pf
Using the relations: $j \partial_{s}=\partial_{t}$ and $j \partial_{t}=-\partial_{s}$, 
the $J$-holomorphic condition (\ref{JHolomorphicCondition}) is equivalent to 
$du(\partial_s)+Jdu(\partial_{t})=0.$ If $u \in \cal{F}$, then $du=d\pi \circ df$ and 
$df_x: \partial_{s} \mapsto (X_A)_{f(x)}$. Then the condition requires that 
$df_x( \partial_t) = J df(\partial_s)=J(X_A)_{f(x)}$.  
Thus, if $u \in \mathcal{F}$,  then $X_B=JX_A$, which implies that $X_b =JX_a$.\\ 
\endpf

In particular, for a map $u \in \mathcal{F}$, using the vectors (\ref{vectors}) to write the 
$J$-holomorphic condition,\cite{Sanborn}
 \begin{equation}
 \bar{\partial}_{J}u=\frac{1}{2}\left( \begin{array}{cc}
v_{1}-w_{2} & w_{1}+v_{2} \\
v_{2}+w_{1} & w_{2}-v_{1} \\
v_{3}-w_{4} & w_{3}+v_{4} \\
v_{4}+w_{3} & w_{4}-v_{3} \\
\vdots & \vdots
 \end{array} \right)=0,
 \end{equation}
which is equivalent to the component-wise Cauchy-Riemann equations:
 \begin{eqnarray}
 v_{\alpha}&=&\frac{\partial f_{\alpha }}{\partial s}=\frac{\partial
 f_{\alpha +1} }{\partial t}=w_{\alpha +1} \nonumber \\
 v_{\alpha +1}&=& \frac{\partial f_{\alpha +1}}{\partial s}=-\frac{\partial
 f_{\alpha }}{\partial t}=-w_{\alpha }.
 \end{eqnarray}
 Thus, the Robertson-Schr\"{o}dinger uncertainty
relation can be interpreted as a comparison of the Riemannian metric area of 
the parallelogram formed by the projected vectors $(X_a)_{[\psi]}$
and $(X_b)_{[\psi]}$ in $T_{[\psi]}(P(\calH))$
to the invariant symplectic area $u^{*}(\Omega)$.
 The metric area of the projection depends on the relative directions of the vectors 
$\hat{A}\psi$ and $\hat{B}\psi$ in $\calH$. For each $p=[\psi] \in P(\calH)$, $\Sigma$ acts as a chart 
domain for the two-dimensional real subspace of $T_p(P(\calH))$ spanned by $\{X_a(p),X_b(p)\}$, which is 
isomorphic to the
subspace of $T_{\psi}(\calH -\{0\})$ spanned by $\{(X_{A}^{horiz})_{\psi},(X_{B}^{horiz})_{\psi}\}$.
The image $u(\Sigma)$ of a map $u \in \mathcal{F}$ is a two-dimensional real submanifold in $P(\calH)$.  
When the map $u$ is $J$-holomorphic, $u(\Sigma)$ is a complex submanifold of $P(\calH)$, that is,
$T_{p}(u(\Sigma))$ is a complex subspace of $T_{p}(P(\calH))$ and, it is in this case that
the uncertainty inequality is saturated. 
This conclusion is consistent with the well-known result that every complex submanifold  of a K\"{a}hler
 manifold is volume minimizing in its homology class.\cite{Lawson}
Observe that the off-diagonal element $C(\hat{A},\hat{B})$
 of the covariance tensor is real, and hence vanishes when $v=-Jw$. 
 \begin{cor}\cite{Sanborn}
If $u \in \mathcal{F}$ is $J$-holomorphic, then the
covariance tensor has minimum determinant and vanishing off-diagonal components 
$C(\hat{A},\hat{B})_{\psi}$. In this case, 
the Robertson-Schr\"{o}dinger uncertainty inequality is saturated.
\end{cor}

 \section{Discussion}
 
A goal for future work is to establish the integral version of the energy identity for$J$-holomorphic  
maps from a specific compact Riemann surface (possibly with boundary) into the quantum state space, and to 
 interpret its physical meaning. It is notable that the covariance tensor can be viewed both as a quantum 
 Fisher information metric and as the Hessian matrix for the energy density function
for a map into state space, suggesting an investigation into the relation between quantum 
information and dynamics on $\Sigma$ by using Hamilton's equations. Also, it would be interesting to 
consider whether the K\"{a}hler property of the quantum state space could be relaxed so that the almost 
complex structure might be non-constant, or even non-integrable and merely compatible with a symplectic 
form. 

{\em Acknowledgement} The author wishes to thank Edwin Ihrig for many helpful discussions about this work.

\end{document}